\documentclass[epj,nopacs]{svjour}

\usepackage{ifthen} 
\newboolean{pdflatex}
\setboolean{pdflatex}{true} 

\newboolean{articletitles}
\setboolean{articletitles}{true} 

\newboolean{uprightparticles}
\setboolean{uprightparticles}{false} 

\def\paperauthors{LHCb collaboration} 
\def\paperasciititle{GPU based optical photon simulation for the LHCb RICH 1 Detector} 
\def\paperkeywords{{High Energy Physics}, {LHCb}} 
\def\papercopyright{\the\year\ CERN for the benefit of the LHCb collaboration} 

\def\paperlicenceurl{https://creativecommons.org/licenses/by/4.0/}

\usepackage{placeins}

\usepackage[top=1in, bottom=1.25in, left=1in, right=1in]{geometry}

\usepackage{microtype}
\usepackage{lineno}  
\usepackage{xspace} 
\usepackage{caption} 

\usepackage{graphicx}  
\usepackage{color}
\usepackage{colortbl}
\graphicspath{{./figs/}} 

\usepackage{amsmath} 
\usepackage{amssymb}
\usepackage{amsfonts}
\usepackage{upgreek} 

\newcommand*\patchAmsMathEnvironmentForLineno[1]{%
\expandafter\let\csname old#1\expandafter\endcsname\csname #1\endcsname
\expandafter\let\csname oldend#1\expandafter\endcsname\csname
end#1\endcsname
 \renewenvironment{#1}%
   {\linenomath\csname old#1\endcsname}%
   {\csname oldend#1\endcsname\endlinenomath}%
}
\newcommand*\patchBothAmsMathEnvironmentsForLineno[1]{%
  \patchAmsMathEnvironmentForLineno{#1}%
  \patchAmsMathEnvironmentForLineno{#1*}%
}
\AtBeginDocument{%
\patchBothAmsMathEnvironmentsForLineno{equation}%
\patchBothAmsMathEnvironmentsForLineno{align}%
\patchBothAmsMathEnvironmentsForLineno{flalign}%
\patchBothAmsMathEnvironmentsForLineno{alignat}%
\patchBothAmsMathEnvironmentsForLineno{gather}%
\patchBothAmsMathEnvironmentsForLineno{multline}%
\patchBothAmsMathEnvironmentsForLineno{eqnarray}%
}

\usepackage{hyperxmp}

\usepackage[pdftex,
            pdfauthor={\paperauthors},
            pdftitle={\paperasciititle},
            pdfkeywords={\paperkeywords},
            pdfcopyright={Copyright (C) \papercopyright},
            pdflicenseurl={\paperlicenceurl}]{hyperref}

\usepackage[colorinlistoftodos,textsize=scriptsize]{todonotes}

\usepackage[bottom,flushmargin,hang,multiple]{footmisc}

\usepackage[all]{hypcap} 

\usepackage{xspace} 
\usepackage{upgreek}

\def\lhcb   {\mbox{LHCb}\xspace}

\def\lhc    {\mbox{LHC}\xspace}

\def\rich   {RICH\xspace}
\def\richone {RICH1\xspace}
\def\richtwo {RICH2\xspace}

\def\MagUp {\mbox{\em Mag\kern -0.05em Up}\xspace}

\ifthenelse{\boolean{uprightparticles}}%
{

 \def\PDelta      {\ensuremath{\Delta}\xspace}                 
 \def\PXi         {\ensuremath{\Xi}\xspace}                 
 \def\PLambda     {\ensuremath{\Lambda}\xspace}                 
 \def\PSigma      {\ensuremath{\Sigma}\xspace}                 
 \def\POmega      {\ensuremath{\Omega}\xspace}                 
 \def\PUpsilon    {\ensuremath{\Upsilon}\xspace}
 \let\oldPi\Pi
 \def\PPi         {\ensuremath{\oldPi}\xspace}

 \def\PB      {\ensuremath{\mathrm{B}}\xspace}                 
                  
 \def\PD      {\ensuremath{\mathrm{D}}\xspace}

 \def\PK      {\ensuremath{\mathrm{K}}\xspace}

 \def\Pi      {\ensuremath{\mathrm{i}}\xspace}

 \def\Ps      {\ensuremath{\mathrm{s}}\xspace}

 \def\thebaroffset{0.0em}
}
{

 \mathchardef\PDelta="7101
 \mathchardef\PXi="7104
 \mathchardef\PLambda="7103
 \mathchardef\PSigma="7106
 \mathchardef\POmega="710A
 \mathchardef\PUpsilon="7107
 \mathchardef\PPi="7105
                  
 \def\PB      {\ensuremath{B}\xspace}                 
                  
 \def\PD      {\ensuremath{D}\xspace}

 \def\PK      {\ensuremath{K}\xspace}

 \def\Pi      {\ensuremath{i}\xspace}

 \def\Ps      {\ensuremath{s}\xspace}

 \def\thebaroffset{0.18em}
}
\newcommand{\offsetoverline}[2][\thebaroffset]{\kern #1\overline{\kern -#1 #2}}%

\makeatletter
\ifcase \@ptsize \relax
  \newcommand{\miniscule}{\@setfontsize\miniscule{4}{5}}
\or
  \newcommand{\miniscule}{\@setfontsize\miniscule{5}{6}}
\or
  \newcommand{\miniscule}{\@setfontsize\miniscule{5}{6}}
\fi
\makeatother

\DeclareRobustCommand{\optbar}[1]{\shortstack{{\miniscule (\rule[.5ex]{1.25em}{.18mm})}
  \\ [-.7ex] $#1$}}

\def\squark    {{\ensuremath{\Ps}}\xspace}

\def\KorKbar {\kern \thebaroffset\optbar{\kern -\thebaroffset \PK}{}\xspace}

\def\D       {{\ensuremath{\PD}}\xspace}

\def\DorDbar {\kern \thebaroffset\optbar{\kern -\thebaroffset \PD}\xspace}

\def\Dp      {{\ensuremath{\D^+}}\xspace}
\def\Dm      {{\ensuremath{\D^-}}\xspace}

\def\DpDm    {\ensuremath{\Dp {\kern -0.16em \Dm}}\xspace}

\def\B       {{\ensuremath{\PB}}\xspace}

\def\BorBbar {\kern \thebaroffset\optbar{\kern -\thebaroffset \PB}\xspace}

\def\Bd      {{\ensuremath{\B^0}}\xspace}

\def\BdorBdbar {\kern \thebaroffset\optbar{\kern -\thebaroffset \Bd}\xspace}

\def\Bs      {{\ensuremath{\B^0_\squark}}\xspace}

\def\BsorBsbar {\kern \thebaroffset\optbar{\kern -\thebaroffset \Bs}\xspace}

\def\Y#1S{\ensuremath{\PUpsilon{(#1S)}}\xspace}

\def\LorLbar     {\kern \thebaroffset\optbar{\kern -\thebaroffset \PLambda}\xspace}

\def\AT#1     {\ensuremath{A_{\mathrm{T}}^{#1}}\xspace}           

\def\C#1      {\ensuremath{\mathcal{C}_{#1}}\xspace}                       
\def\Cp#1     {\ensuremath{\mathcal{C}_{#1}^{'}}\xspace}                    
\def\Ceff#1   {\ensuremath{\mathcal{C}_{#1}^{\mathrm{(eff)}}}\xspace}        
\def\Cpeff#1  {\ensuremath{\mathcal{C}_{#1}^{'\mathrm{(eff)}}}\xspace}       
\def\Ope#1    {\ensuremath{\mathcal{O}_{#1}}\xspace}                       
\def\Opep#1   {\ensuremath{\mathcal{O}_{#1}^{'}}\xspace}                    

\newcommand{\aunit}[1]{\ensuremath{\text{\,#1}}}       

\newcommand{\tev}{\aunit{Te\kern -0.1em V}\xspace}
\newcommand{\gev}{\aunit{Ge\kern -0.1em V}\xspace}
\newcommand{\mev}{\aunit{Me\kern -0.1em V}\xspace}
\newcommand{\kev}{\aunit{ke\kern -0.1em V}\xspace}
\newcommand{\ev}{\aunit{e\kern -0.1em V}\xspace}
 
\newcommand{\mevc}{\ensuremath{\aunit{Me\kern -0.1em V\!/}c}\xspace}
\newcommand{\gevc}{\ensuremath{\aunit{Ge\kern -0.1em V\!/}c}\xspace}
\newcommand{\mevcc}{\ensuremath{\aunit{Me\kern -0.1em V\!/}c^2}\xspace}
\newcommand{\gevcc}{\ensuremath{\aunit{Ge\kern -0.1em V\!/}c^2}\xspace}

\def\mm   {\aunit{mm}\xspace}

\def\ghz  {\ensuremath{\aunit{GHz}}\xspace}

\def\gsim{{~\raise.15em\hbox{$>$}\kern-.85em
          \lower.35em\hbox{$\sim$}~}\xspace}
\def\lsim{{~\raise.15em\hbox{$<$}\kern-.85em
          \lower.35em\hbox{$\sim$}~}\xspace}

\def\mrad{\aunit{mrad}\xspace}

\def\gauss      {\mbox{\textsc{Gauss}}\xspace}
\def\geant      {\mbox{\textsc{Geant4}}\xspace}

\def\opticks      {\mbox{\textsc{Opticks}}\xspace}
\def\optix      {\mbox{\textsc{OptiX}}\xspace}

\def\tell1  {TELL1\xspace}
\def\ukl1   {UKL1\xspace}

\def\cfourften     {\ensuremath{\mathrm{ C_4 F_{10}}}\xspace}
\def\cffour        {\ensuremath{\mathrm{ CF_4}}\xspace}

\newcommand{\lhcborcid}[1]{\href{https://orcid.org/#1}{\hspace*{0.1em}\raisebox{-0.45ex}{\includegraphics[width=1em]{figs/orcidIcon.pdf}}}}

\usepackage{cite} 
\usepackage{mciteplus}

\usepackage{longtable} 

\title{GPU-based optical photon simulation for the LHCb~RICH~1 Detector}
\abstract{We present the investigation of the use of \opticks, a GPU-accelerated optical photon interface with the \lhcb detector simulation, to improve computation time of optical photon propagation. The hybrid workflow, combining the particle simulation package \geant and \opticks, offloads optical photon propagation to GPUs, thereby accelerating the overall simulation process. The consistency of the results obtained from \geant and \opticks simulations is verified with a simplified \lhcb \richone detector geometry, demonstrating the feasibility of the proposed approach. In addition, the ongoing transition to the NVIDIA \optix 7 API and re-structuring of \opticks code is discussed within the context of HEP simulation workflows, with caveats explored.
}

\author{Yunlong Li\inst{1}\and Adam Davis\inst{1}\and Sajan Easo\inst{2}\and Keith Evans\inst{1}\and  Evelina Mihova Gersabeck\inst{1}\and Marco Gersabeck\inst{1}\and Lucas George Girardey\inst{2,3}\and Raja Nandakumar\inst{2}\and Annabelle Jane Pointer\inst{3}}
\institute{School of Physics and Astronomy, University of Manchester, M13 9PL, Manchester, United Kingdom\and Rutherford Appleton Laboratory, Harwell Campus, Didcot, OX110QX, United Kingdom \and Southampton University, Southampton, United Kingdom}

\begin{document}


\maketitle{}







\section{Introduction}
\label{sec:introduction_Opticks}

\begin{figure}[!htb]
    \begin{center}
    \includegraphics[width=0.99\columnwidth]{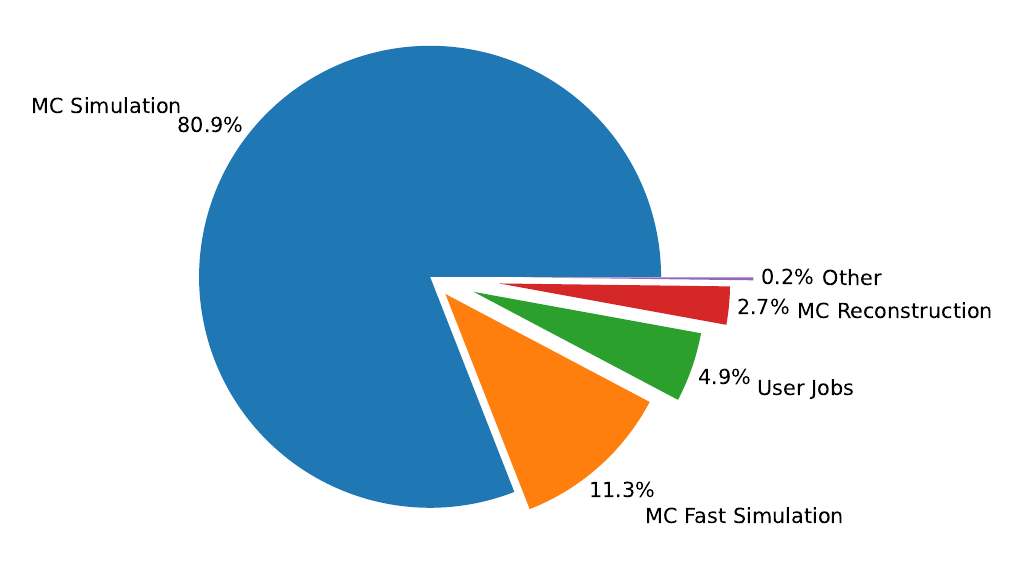}
    \vspace{-3em}
    \end{center}
    \caption{ LHCb experiment CPU usage broken down by task from 2022 to present. As illustrated in blue and orange, the dominant CPU user is the simulation of collisions. These numbers are from productions of simulated event samples for \lhcb utilising the DIRAC project\cite{Mathe_2015} and monitored via the DIRAC web portal \cite{Stagni_2017}. The largest chunk, labelled MC Simulation, represents the generation and propagation of particles throughout the complete detector geometry as described in \geant. The slice labelled MC Fast Simulation is similar to MC Simulation, but utilises the ReDecay framework\cite{LHCb-DP-2018-004}. Finally, MC Reconstruction represents the pattern recognition and track reconstruction, particle combinations and application of hardware and software level triggers. Jobs labelled as User are individual submissions, with the final slice representing all other contributions. Similar dominance of MC simulation is seen during past data-taking periods.
    }
    \label{fig:CPU_usage_forecast}
    \vspace{-2em}
\end{figure}

Simulation of proton-proton collisions dominates the computing resources of the \lhcb experiment~\cite{LHCb_2008}. This dominance is illustrated in Figure~\ref{fig:CPU_usage_forecast}, which presents the CPU usage at \lhcb from January 2022 to present\cite{Mathe_2015,Stagni_2017}. The CPU usage is dominated by detailed simulations based on \geant~\cite{Agostinelli:2002hh,Allison:2006ve}, with a secondary contribution from fast simulation options, such as ReDecay\cite{LHCb-DP-2018-004} and partial detector simulation. Through explicit measurements, it is known that the majority of time in \lhcb simulation is dominated by the propagation of particles through the detector within the \geant framework\cite{b_coutri_sim_talk}. 

\begin{figure}[!thb]
    \begin{center}
    \includegraphics[width=0.99\columnwidth]{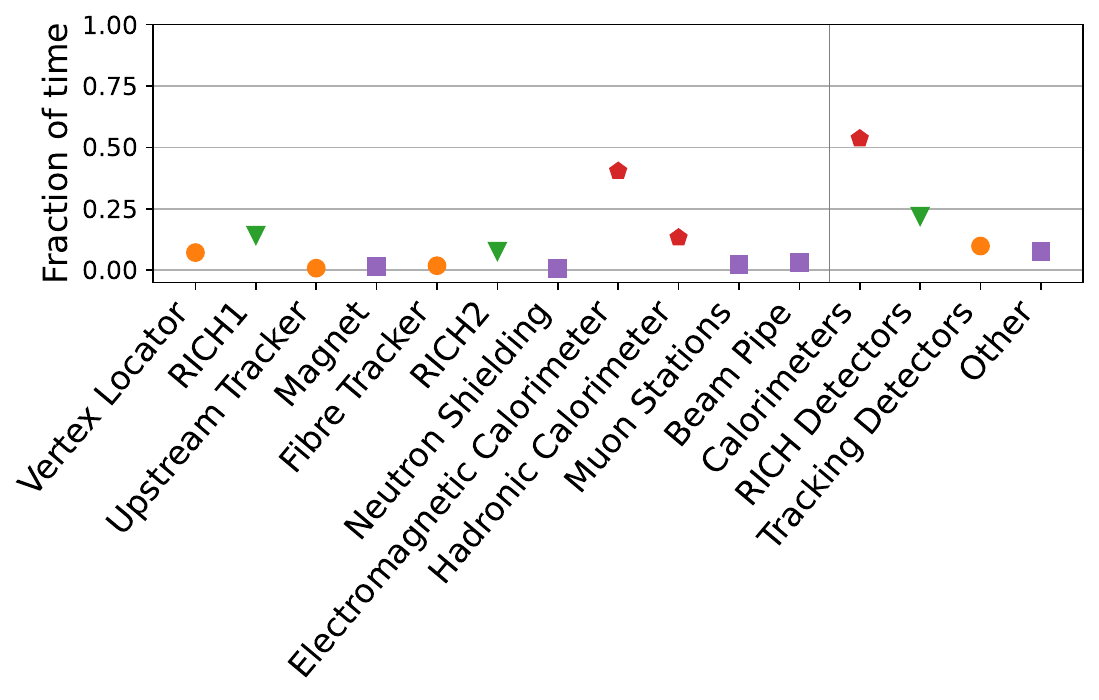}
    \end{center}
    \caption{Time spent in each sub-detector relative to the total time for simulating 100 events, produced by the \lhcb performance and regression (PR) tests. The left part of the plot shows individual sub-detector contributions and the right part shows sums by detector type, indicating that simulation of the RICH detectors amount to about one quarter of the total time. }
    \label{fig:time_spent_subdetector}
    \vspace{-2em}
\end{figure}

The projected use of computing resources will only grow as we pass into the High Luminosity-LHC era, greatly surpassing the resources pledged to the experiment. As a result, research and development into simulation on diverse computing architectures is necessary. Graphical Processing Units (GPUs), are under active investigation, particularly as they have been designed for high-efficiency parallel processing, allowing for considerable computing cost reduction and increase in simulation speed. The \lhcb collaboration utilises \geant for its fully detailed simulation~\cite{Agostinelli:2002hh,Allison:2006ve}. Numerous research and development endeavours have been initiated within the larger High Energy Physics (HEP) community, targeting the execution of specific simulation components on GPUs. Figure~\ref{fig:time_spent_subdetector} illustrates the time used in \geant simulations per sub-detector for the latest release of the \lhcb simulation package \gauss~\cite{LHCb-PROC-2011-006}, highlighting that the computational resources for simulation are dominated by the calorimeters and the two Ring Imaging Cherenkov (\rich) detectors. These two systems are associated with two distinct physical processes: electromagnetic showers and the transportation of optical photons. The \geant collaboration is currently investigating methods to incorporate GPUs for modelling both processes. The {\sc Celeritas}~\cite{doecode_94866} and {\sc ADePT} projects~\cite{Amadio_2023} represent such development initiatives, aiming to simulate the electromagnetic physics utilising a GPU. Furthermore, the \opticks package~\cite{refId0} provides an interface between \geant and the NVIDIA \optix ray tracing engine to simulate photon interactions. This paper reports efforts testing the \opticks package for use by the \lhcb simulation framework, further discussed in Sec.~\ref{subsec:opticks_richone}. \par
The acceleration of optical photon propagation is not only a necessity for future \lhcb upgrades, but also for numerous neutrino experiments, due to the significant presence of cosmic muons, which can generate critical background noise for neutrino detectors such as Daya Bay~\cite{An_2016} and JUNO~\cite{JUNO_2016}. For example, a muon with a typical energy of 200\gev traversing the JUNO scintillator is projected to produce tens of millions of optical photons~\cite{refId0}. Similarly to the \lhcb use case, the propagation of optical photons in these experiments accounts for over 99\% of CPU time and imposes rigorous memory limitations on the storage of each event~\cite{refId0}. Consequently, it is beneficial to explore transfer of optical photon simulation workloads from CPUs to GPUs, utilising potential programs such as \opticks that can perform these actions, and which can therefore accelerate the simulation of any experiment relying on \geant optical photon propagation.

\subsection{\optix ray tracing engine}
\label{subsec:OptiX}
The NVIDIA \optix~\cite{Parker10OptiX} ray tracing engine constitutes a versatile programmable system, specifically devised for NVIDIA GPUs and other highly parallel architectures. As a domain-specific just-in-time compiler, the \optix engine operates on the fundamental premise that the majority of ray tracing algorithms can be executed through a series of programmable operations, encompassing ray generation, material shading, object intersection, and scene traversal. \optix ray tracing pipelines are assembled from a limited set of user-provided programmes, coded utilising CUDA\cite{10.1145/1365490.1365500} --- a parallel computing platform and programming model established by NVIDIA for general-purpose computing on GPUs. This facilitates the implementation of an extensive variety of ray tracing-based algorithms and applications, including interactive rendering, offline rendering, collision detection systems, artificial intelligence, and scientific simulations, such as sound propagation. Spatial index data structures, exemplified by the boundary volume hierarchy (BVH)~\cite{10.1145/1572769.1572771}, primarily serve to expedite intersections between photon paths and geometry structures. \optix offers a flexible interface, appropriate for a broad spectrum of applications, to regulate its acceleration structures and facilitate intersections with any geometric form. The \optix acceleration structure system supports adaptable instancing, which pertains to the low-overhead replication of scene geometry via referencing the same data multiple times without duplication. This enables the acceleration of constructing multiple instances of identical geometry, such as the pixel Multianode Photomultiplier Tubes (MaPMTs) within the \lhcb \richone detector. \par
In August 2019, NVIDIA unveiled \optix 7, superseding the previous \optix 6.5 API and offering a low-level, CUDA-centric API. The \optix 7 API gives explicit control over memory management, compilation, and launches to the application, whilst preserving the ray tracing programming model and shader types. The \opticks package is currently undergoing a transition from the \optix 6.5 API to \optix 7\cite{opticks_chep_2023}. Unless indicated otherwise, the subsequent analyses are performed utilising the \optix 6.5 API.

\subsection{Hybrid \geant and \opticks workflow}
\label{subsec:hybrid_workflow}
To facilitate the offloading of optical photon propagation in \geant to GPUs, while maintaining the simulation of other particles on CPUs, all elements of optical photon generation and propagation are converted into a suitable format and uploaded to the GPU by \opticks. The primary components within the hybrid workflow encompass detector description including material and surface properties, as well as optical photon generation and propagation as determined by optical physics. These aspects are covered in detail in Ref~\cite{refId0}, and are summarised here. \par

\subsubsection{Geometry translation}
\label{subsubsec:geometry_translation}
During the initialisation phase, \geant geometries are initially translated into the corresponding \opticks geometry buffers. This translation encompasses the conversion of materials, surfaces, solids, volumes, and sensors, and construction of complex geometries is accomplished utilising these primitive shapes through constructive solid geometry (CSG) modelling. The user-provided \geant geometry is then translated into a parallel tree of \opticks nodes. Each volume boundary is allocated an index uniquely identifying the combination of four indices representing outer and inner materials and outer and inner surfaces. Outer/inner surfaces manage inward/outward-directed photons, enabling the translation of \geant border and skin surface functionality. Surfaces exhibiting a non-zero efficiency property are employed to identify sensor volumes. To circumvent repetition of geometry translation in each run, the \opticks geometry is cached within a directory structure for future use. \par
Upon invoking the visualisation or actual simulation, the \opticks geometry buffers are further translated into an \optix geometry node graph, which is also utilised to configure the resulting acceleration structures. One such visualisation is presented in  Figure~\ref{fig:rendering_RICH1}, which exhibits a ray traced rendering of the \richone detector geometry.

\subsubsection{Optical photon generation and propagation}
\label{subsubsec:Optical_photon}
We summarise a few key points on optical photon generation and propagation here for future reference. Optical physics processes, such as scattering, absorption, scintillator re-emission, and boundary processes, are implemented in CUDA functions based on the corresponding \geant implementations. Rather than generating photon secondary tracks iteratively, pertinent ``\texttt{genstep}'' parameters, including the number of photons to generate and the line segment along which they should be generated, are collected. These \texttt{gensteps}, in conjunction with CUDA ports of the Cherenkov and scintillation generation, facilitate the direct generation of photons on the GPU within the ray generation programme provided to \optix. Ultimately, only a small proportion of the photons detected as hits on the sensor volumes are copied back to the CPU and incorporated into \geant hit collections. 

\section{Test of \opticks with \lhcb \richone detector}
\label{subsec:opticks_richone}
\begin{figure}[!htb]
    \begin{center}
    \includegraphics[width=0.45\textwidth]{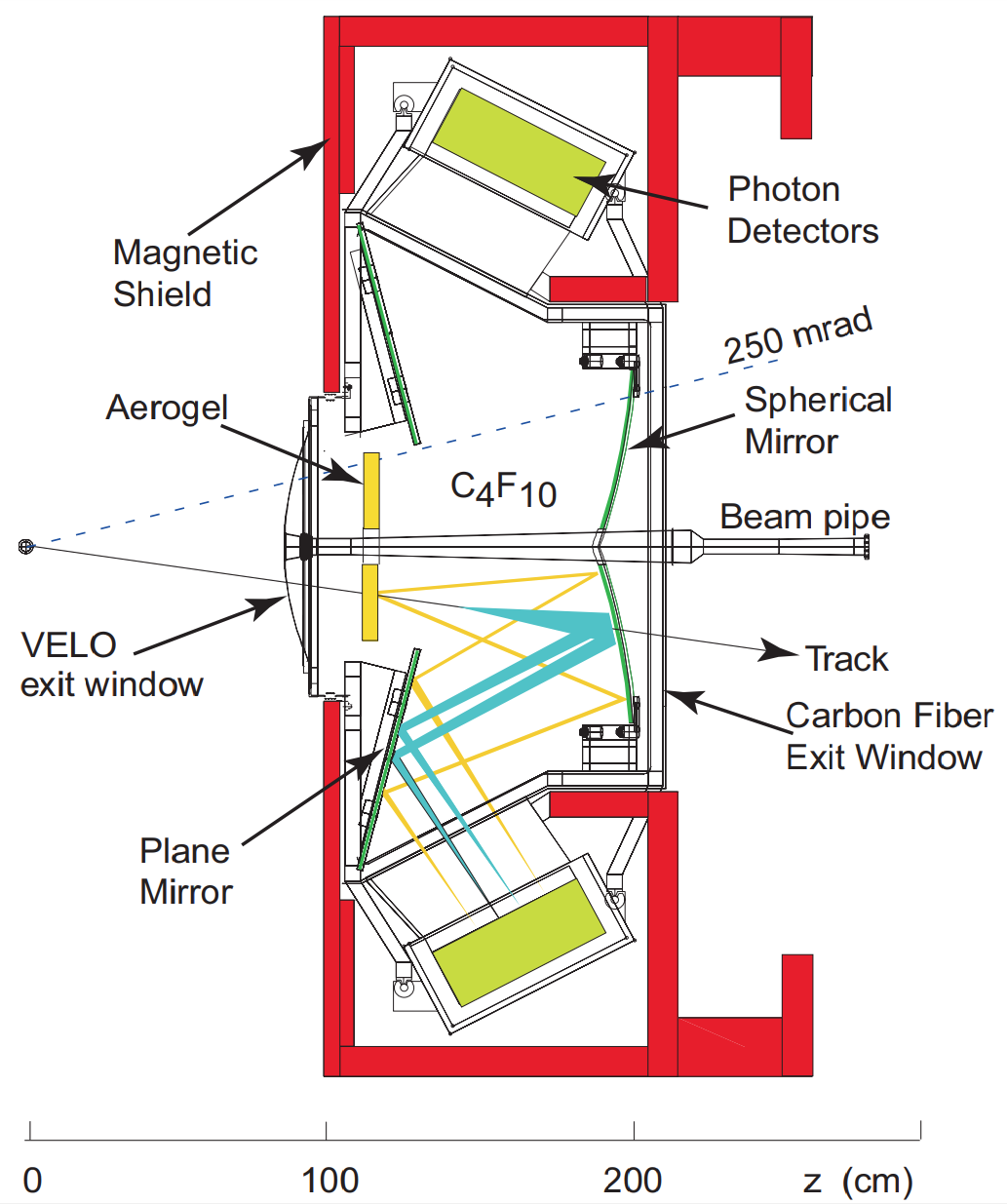}
    \includegraphics[width=0.35\textwidth]{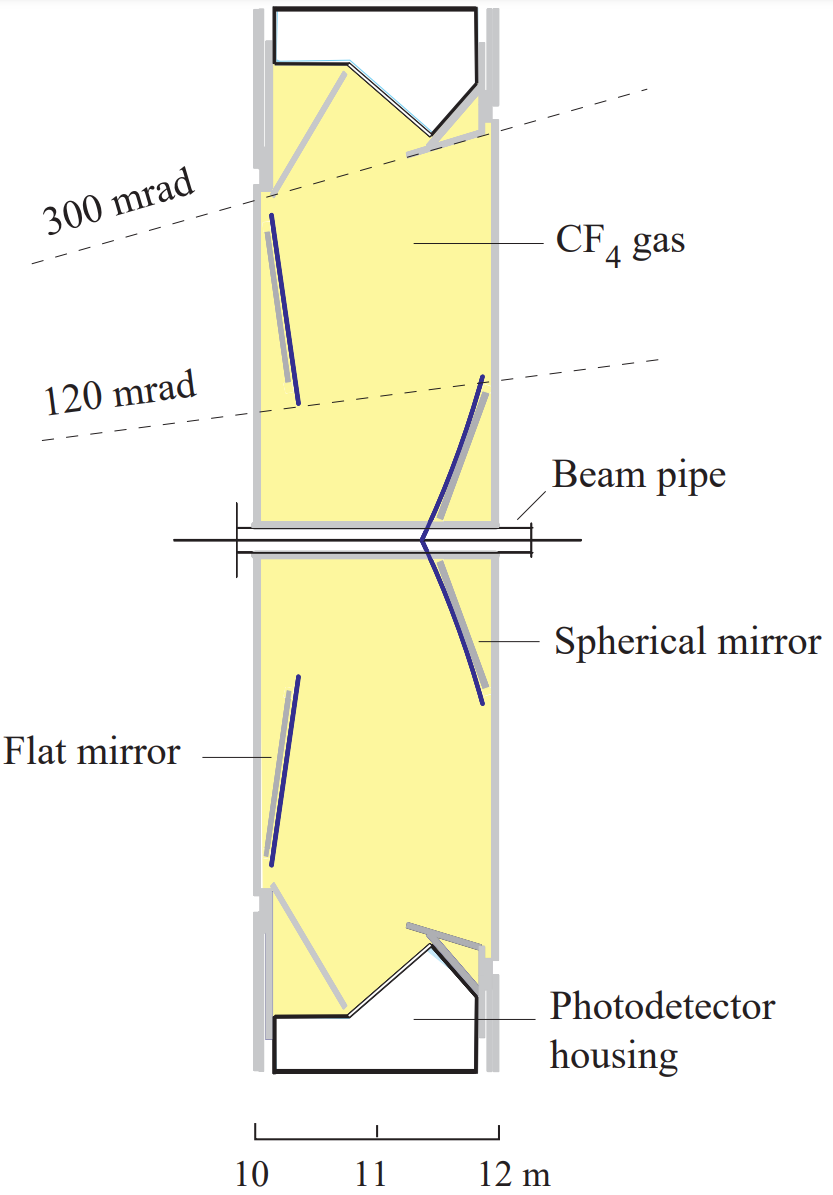}
    \end{center}
    \caption{ Schematic view of \richone (top) and \richtwo (bottom) detectors and their optical systems. In Run 2 of the \lhc, the silica aerogel in the \richone was removed and is not considered further. Taken from~\cite{LHCb_2008,Amato:494263}.}
    \label{fig:RICH}
\end{figure}

The \lhcb detector employs two \rich detectors for particle identification~\cite{Amato:494263,Collaboration:1624074}. These detectors operate on the principle of Cherenkov radiation produced by charged particles traversing the radiator medium at velocities surpassing the speed of light in the radiator. The emitted light forms a conical pattern, characterised by an angle $\theta_c$ contingent upon the particle's velocity, given by
\begin{equation}
    \label{equ:Cherenkov_angle}
    \cos\theta_c = \frac{1}{n\beta},
\end{equation}
where $\beta=v/c$ is the velocity of the particle normalised to the speed of light in the vacuum and $n$ is the refractive index of the medium, which can depend on the wavelength of light traversing the medium. By ascertaining the Cherenkov angle, and combining it with the particle momentum derived from tracking measurements, a mass hypothesis can be tested for the particle, facilitating its identification. For this reason, statistical comparisons between any optical photon propagation offloading and nominal \geant techniques are imperative. 
\lhcb is equipped with a pair of \rich detectors, utilising two distinct radiators: \richone employs a \cfourften radiator, possessing a refractive index of 1.0014, to identify tracks in the momentum range of 2 to 60\gevc, while \richtwo is filled with \cffour gas, with a refractive index of 1.0005, specifically for momenta as high as 100\gevc. Both \rich detectors incorporate dual sets of mirrors. The primary mirrors are spherical and are tilted, reflecting Cherenkov photons towards one of the secondary mirrors. These secondary mirrors are effectively planar, redirecting the photons outside the \lhcb detector acceptance, where the photodetectors are situated (refer to Fig.~\ref{fig:RICH}). This plane aligns with the focal plane of the corresponding segment of the optical system, causing Cherenkov photons produced by the same track to focus into a ring on the photon detection plane.
\vspace{-2em}
\FloatBarrier
The photodetectors selected for \richone consist of MaPMTs, which are structured as matrices of $8\times8$ anodes. \richone is outfitted with 1-inch MaPMT modules, each with a pixel size of $2.88\times2.88\mm^2$. The Cherenkov angle resolution for \richone is expected to be approximately 0.83\mrad for a single photon~\cite{LHCb:2023hlw}.

\subsection{Simplified \richone geometry}
\label{subsec:simplified_rich}
The full \richone system includes support structures, and is at a level of detail which is unnecessary for validation of the optical photon simulation presented here. Instead, we use a simplified \rich geometry to validate the performance of \opticks and check the consistency of \geant simulation and hybrid workflow simulation in a convenient way. This simplified \rich contains only a singular set of spherical and planar mirrors in the upper section of \richone, and it does not incorporate features such as the beam pipe, magnetic shielding, and exit window --- the optical geometry and the radiator are identical to those utilised in the full \richone detector. Detector properties, such as mirror reflectivity and MaPMT quantum efficiency corrections, are identical to those employed in the full \lhcb \richone.

Furthermore, to obtain the \geant and \opticks simulation results simultaneously, a \texttt{Rich\_Simplified} program based on the \opticks \texttt{G4Opticks} class has been developed. This program allows direct comparison and validation, as well as visualisation of the simulation results by propagation of the same event within the two frameworks. 

\subsection{Measurement of Timing and Statistical Comparisons}
The measurement of timing is performed two ways, first using built-in timing capabilities of the \texttt{Rich\_Simplified} program, separately for \geant and \opticks timing, and second, utilising the NVIDIA Nsight tool~\cite{nvidia_nsight}, a package of debugging and profiling tools designed for NVIDIA GPU-accelerated applications.  

\section{Results}
\subsection{Rendering of Geometries}
\begin{figure}[!th]
    \begin{center}
    \includegraphics[width=0.8\columnwidth]{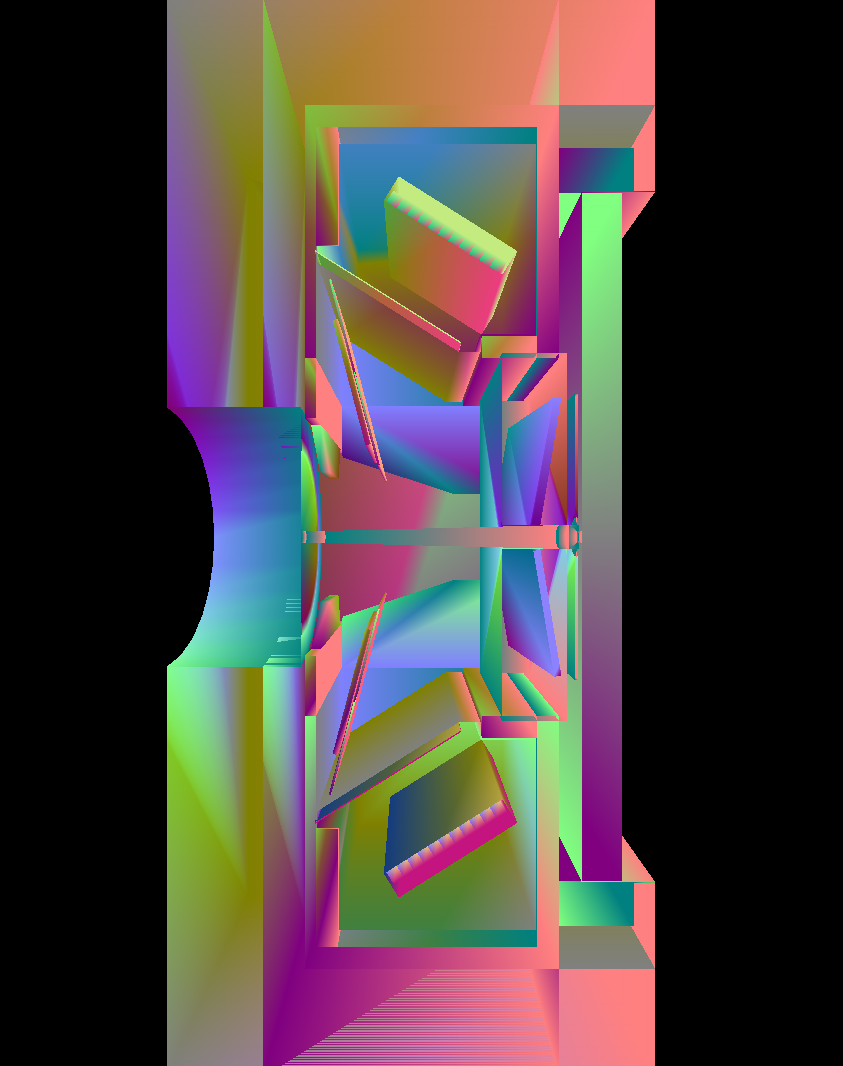}
    \includegraphics[width=0.8\columnwidth]{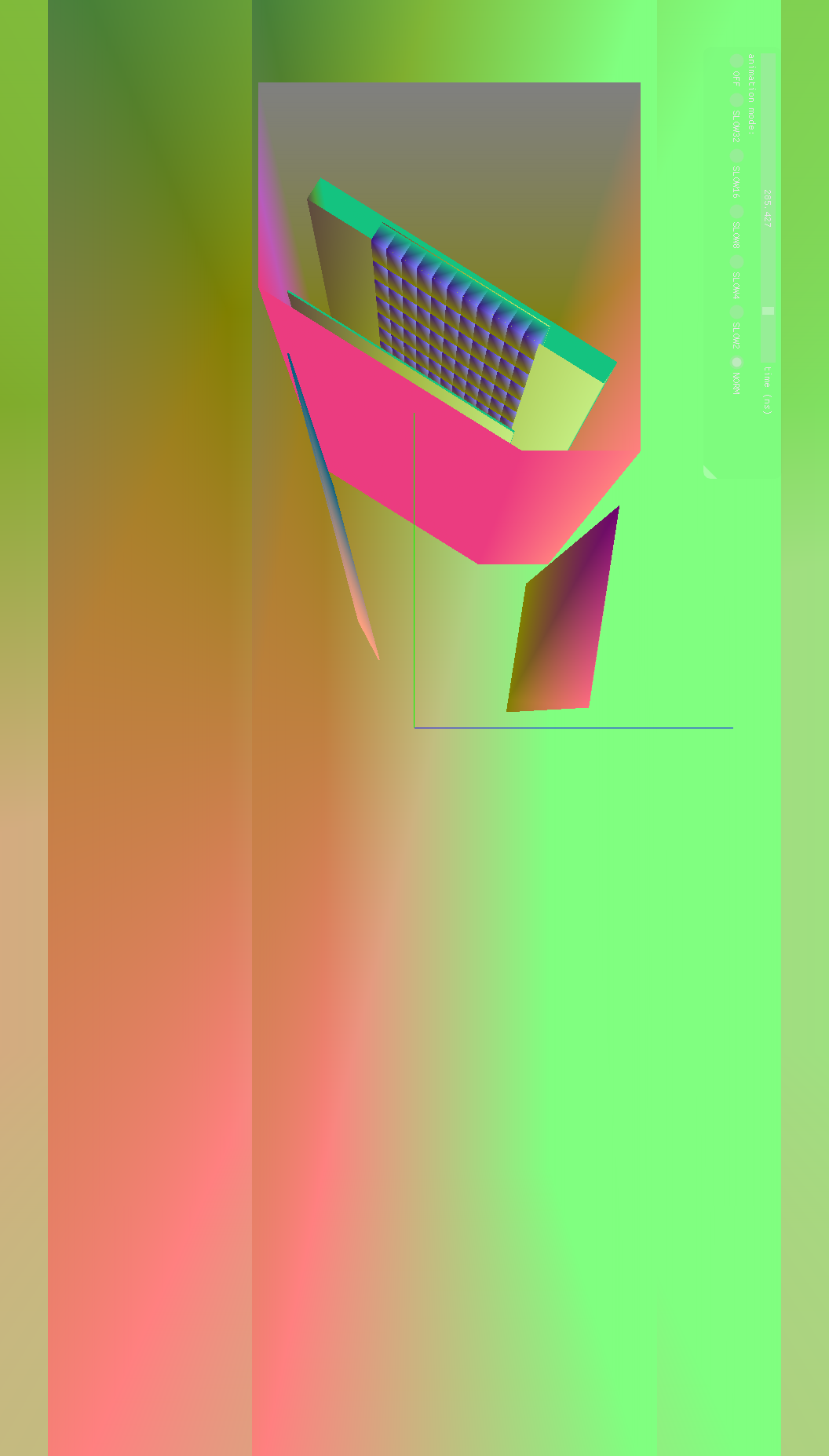}
    \end{center}
    \caption{OpenGL rendering of the full \richone (top) and simplified \richone (bottom) geometry. Both are converted from \geant into an \opticks geometry. The geometry is cut in the centre to allow the visualisation of components. The colour gradient is artificial and implemented to display the surfaces. }
    \label{fig:rendering_RICH1}
    \vspace{-2em}
\end{figure}

The incorporation of the \richone and \texttt{Rich\_Simplified} geometries are shown in Figure~\ref{fig:rendering_RICH1}. These renders show the capability of incorporation of arbitrary geometries within the \opticks framework. However, several iterations were necessary before an adequate geometry representation was achieved. These include overlapping volumes as seen by \opticks, incompatible boolean subtractions between \geant based geometries, and incompatibilities between detection planes of \geant and \opticks were discovered and subsequently updated in the geometry itself. These issues were reported to core developers.
\FloatBarrier
\subsection{Timing and Profiling}
\begin{figure}[!th]
    \begin{center}
    \includegraphics[width=0.99\columnwidth]{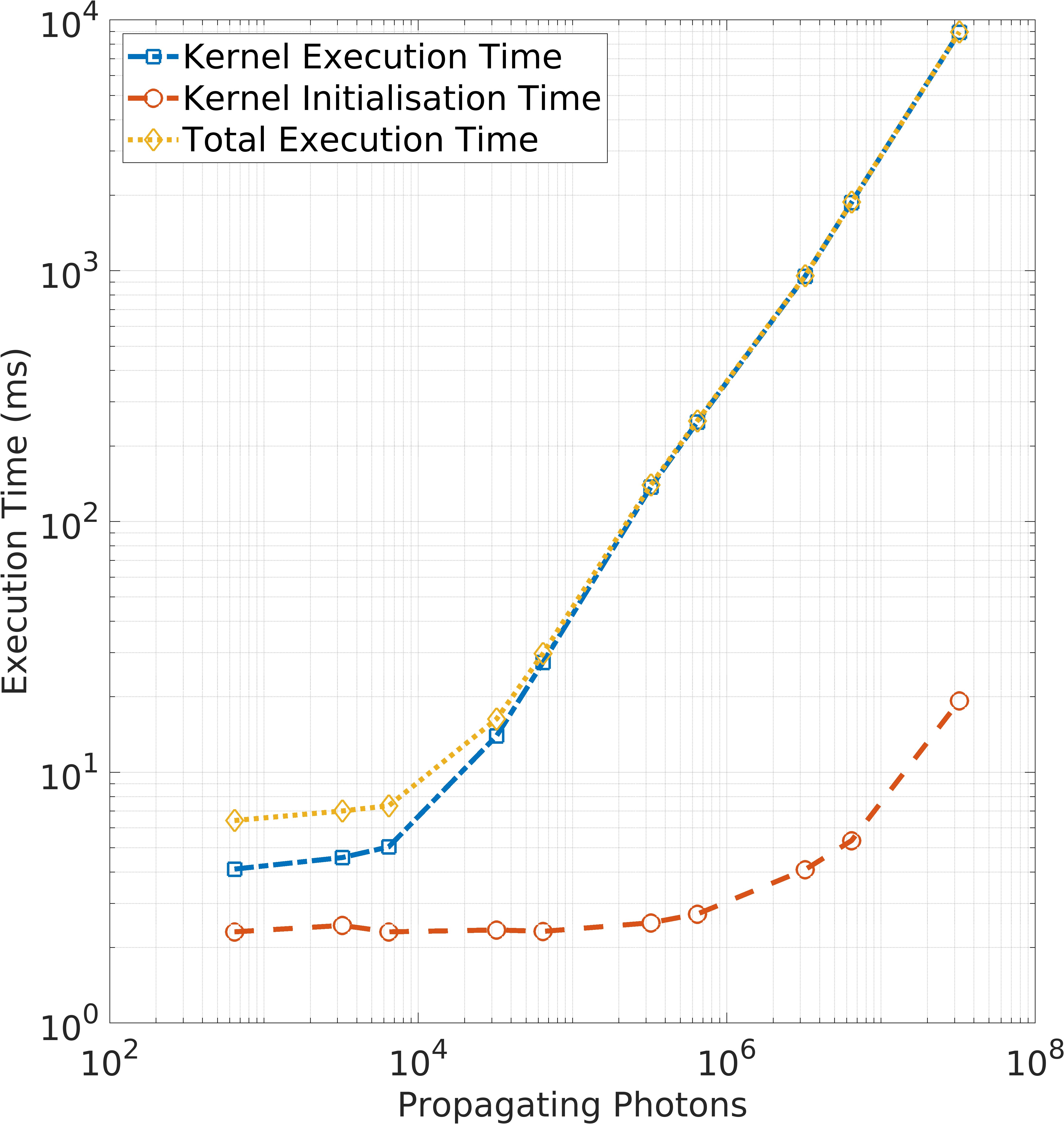}
    \includegraphics[width=0.99\columnwidth]{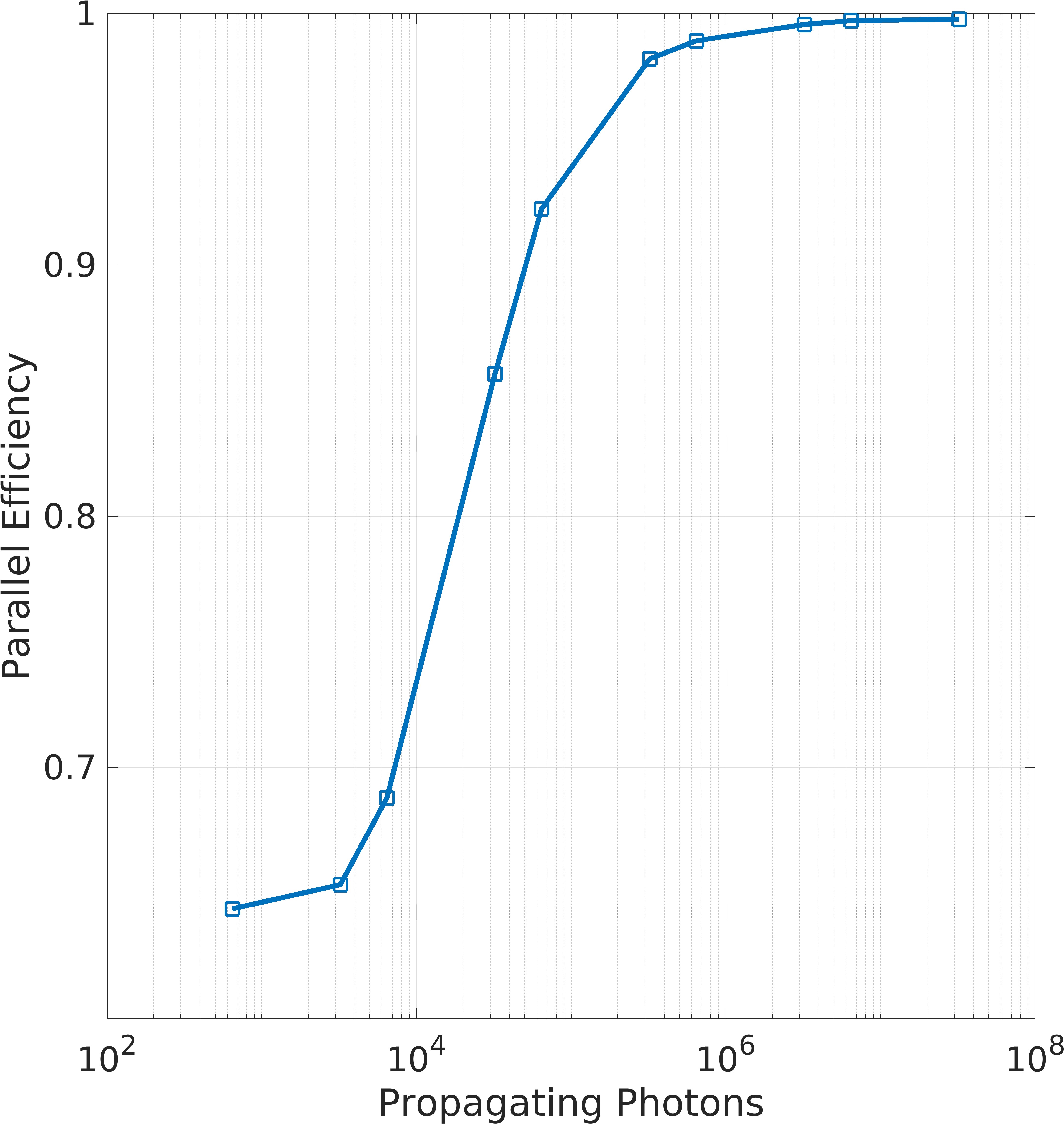}
    \end{center}
    \caption{The top figure illustrates the kernel initialisation time (represented by the red line), photon propagation time (depicted by the blue line), and total CUDA kernel execution time (shown by the yellow line) as the initial photon numbers increase. The bottom figure displays the kernel execution efficiency, defined as the CUDA kernel execution time divided by the total compute time, as a function of the initial photon numbers.}
    \label{fig:kernel_time}
    \vspace{-2em}
\end{figure}

For timing studies we compared the performance using an Intel(R) Xeon(R) Silver 4210R 2.4\ghz CPU and two NVIDIA Tesla T4 GPUs. The timing results of the \texttt{Rich\_Simplified} programme with \opticks and \geant are reported in Figure~\ref{fig:kernel_time}. Reported CPU propagation times are measured using a single core, and GPU propagation times are presented per GPU used, with two GPUs used in total. 
Figure~\ref{fig:kernel_time} top shows that there is an underlying overhead of device initialisation within the \opticks code and is verified via the NVIDIA Nsight tool. This time corresponds to 35\% of the total time of the program for a single muon propagated through the Simplified RICH geometry. This effect is further quantified by Figure~\ref{fig:kernel_time} bottom, which shows the photon propagation time divided by the total kernel execution time, indicating a measure of efficiency of the offload process.  We find that for the \texttt{Rich\_Simplified} program, one needs to propagate $10^5$ photons per kernel execution to ensure at least 95\% of the execution time is used to propagate photons.

\subsection{Statistical Comparisons}
\begin{figure}[!ht]
    \begin{center}
    \includegraphics[width=0.45\textwidth]{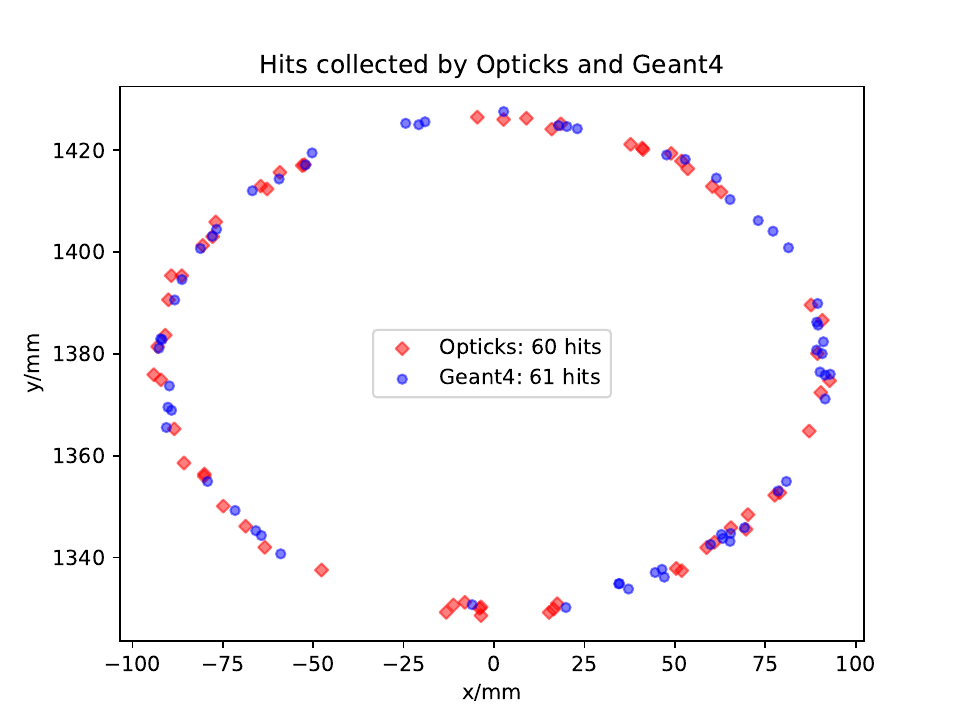}
    \includegraphics[width=0.45\textwidth]{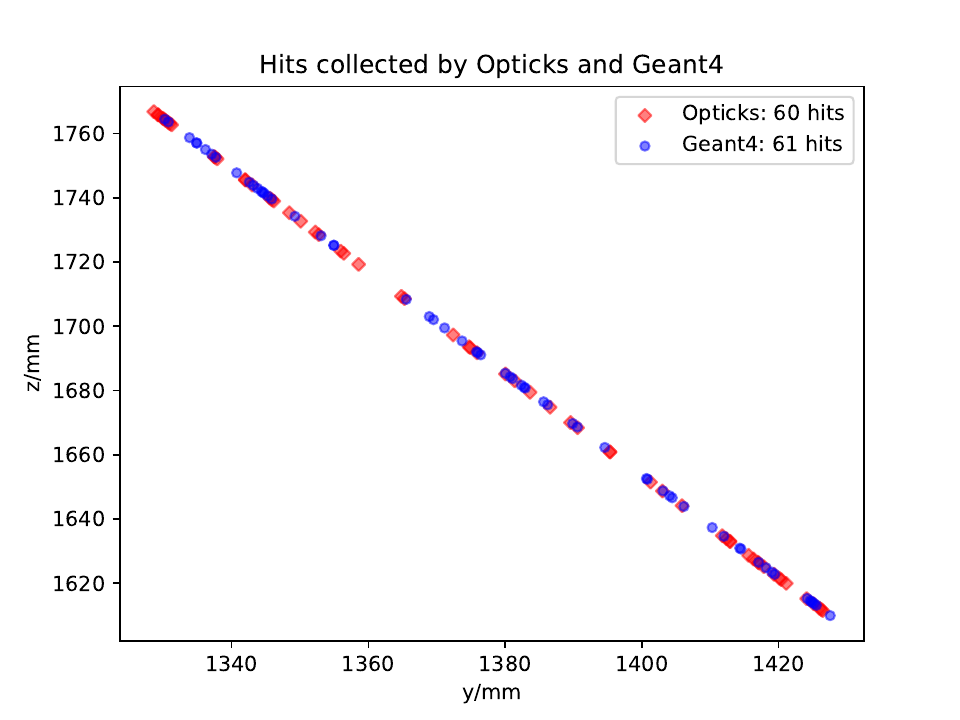}
    \end{center}
    \caption{The positions of the hits on the MaPMTs' photocathode are obtained from the simulation results of \geant (in blue) and \opticks (in red). The top figure displays the positions of the hits in the $x-y$ plane, while the bottom figure shows the positions in the $y-z$ plane.}
    \label{fig:hits_positions}
    \vspace{-2em}
\end{figure}

\begin{figure}[!ht]
    \begin{center}
    \includegraphics[width=0.45\textwidth]{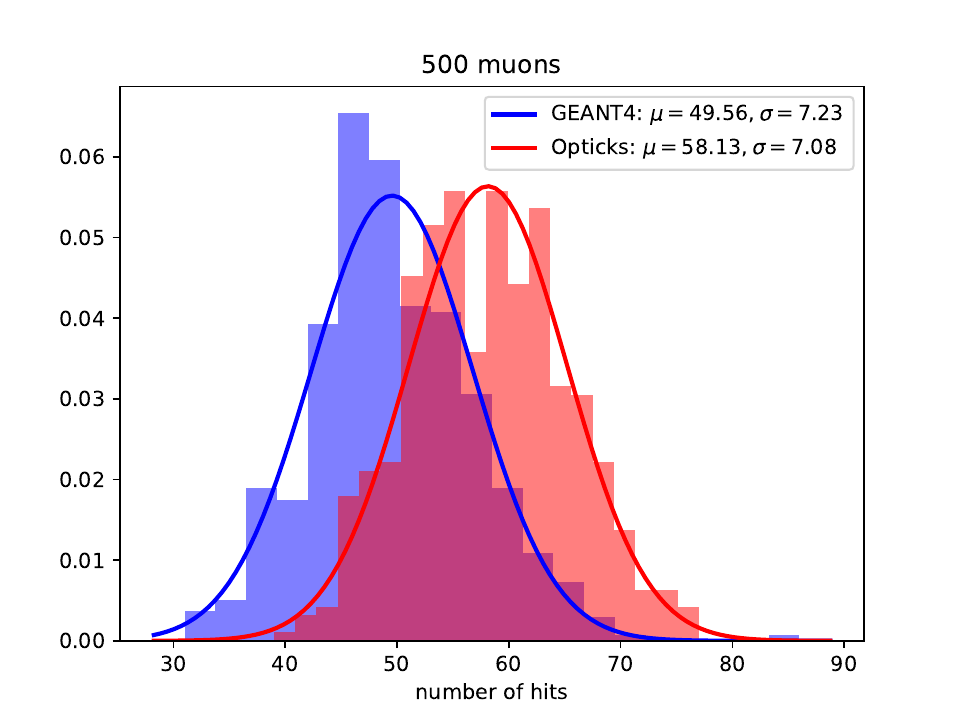}
    \includegraphics[width=0.45\textwidth]{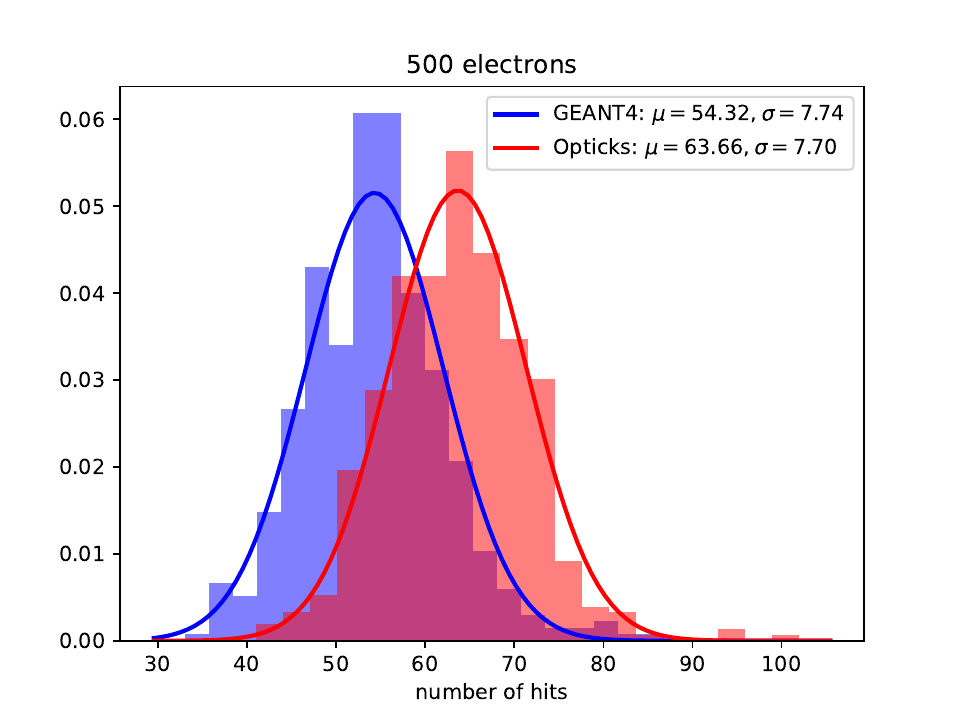}
    \end{center}
    \caption{The top figure displays the distributions of hit numbers from 500 muons at 10\gev, simulated by \opticks (in red) and \geant (in blue). The means and standard deviations obtained from Gaussian function fits are indicated on the plots. The bottom figure presents the distributions of hit numbers from 500 electrons at 10\gev. }
    \label{fig:stat_500events}
    \vspace{-2em}
\end{figure}

The agreement between \geant and \opticks simulation results is checked by running the \texttt{Rich\_Simplified} program and performing a statistical comparison of the results. Test hardware is a single NVIDIA Tesla T4 GPU with Intel(R) Xeon(R) Silver 4210R 2.40\ghz, 62G CPUs. Figure~\ref{fig:hits_positions} displays the hits collected on the MaPMT photocathodes, obtained from a muon passing through the simplified \richone with an energy of 10\gev. It is worth noting that these hits are acquired with accounting for quantum efficiency corrections. Upon the inclusion of quantum efficiency corrections, both \geant and \opticks simulation results yield approximately 60 hits, consistent in position of the detection plane and also yielding a consistent Cherenkov ring. Figure~\ref{fig:stat_500events} displays the distributions of hit numbers gathered from 500 events by both \geant and \opticks simulations, fitted using Gaussian functions. The mean values of collected hits are consistent within one standard deviation, and both simulations demonstrate a similar Gaussian distribution, suggesting accurate translations of both geometry and physical processes. The variance in the mean values stems from two individual points:  first, \geant allows the possibility for user defined physics processes to be implemented, which is taken advantage of in the \texttt{Rich\_Simplified} program, for optimisations such as the elimination multiple bounces of trapped optical photons. The \opticks package, however, has its own set of physics processes which have been incorporated within the library. These differing physics processes account for the shift in the mean position of the peaks. This difference is validated by removing certain user-implemented physics processes, resulting in a notable alteration in the observed discrepancies.

\begin{figure}[!ht]
    \begin{center}
    \includegraphics[width=0.45\textwidth]{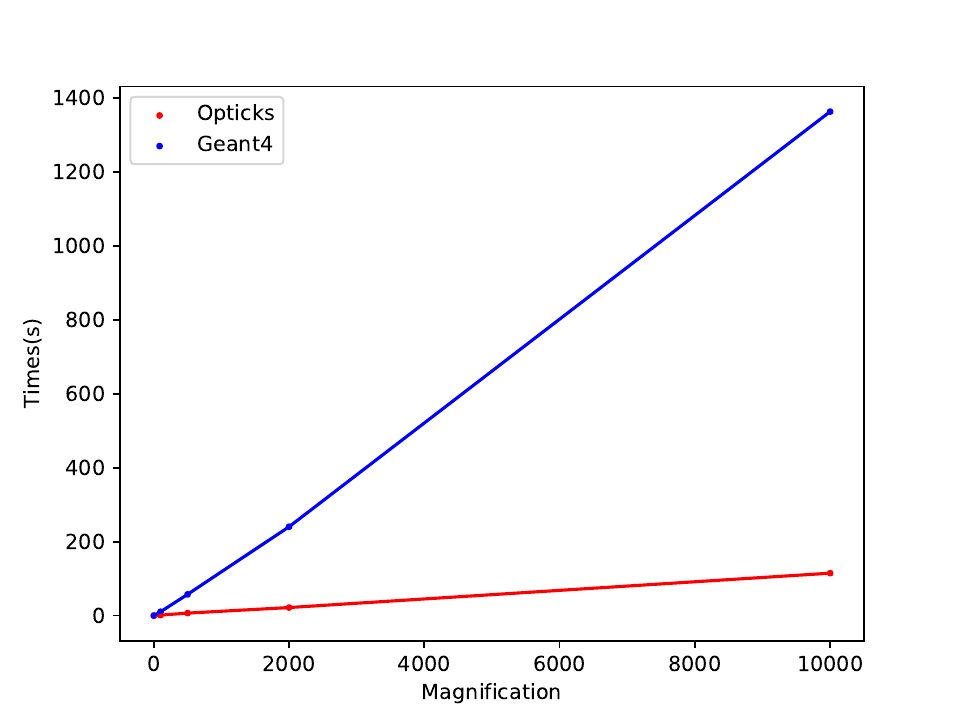}
    \includegraphics[width=0.45\textwidth]{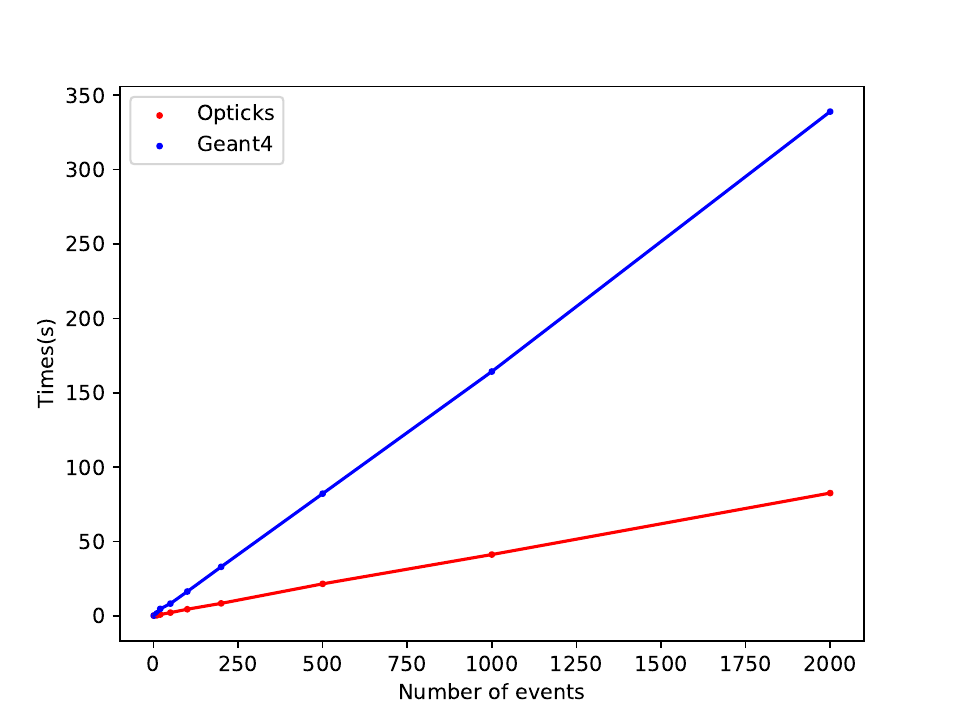}
    \end{center}
    \caption{The computational times required by \geant (in blue) and \opticks (in red) simulations are displayed in the figures. The top figure illustrates the times as the number of photons generated in a single event is increased by a magnification factor, while the bottom figure depicts the times as the number of events is increased. }
    \label{fig:speed-up}
    \vspace{-2em}
\end{figure}

Lastly, the efficiency improvement of \opticks simulation over pure \geant simulation is evaluated in two ways: by running a single event with an increased number of generated photons, achieved by multiplying the expected number of generated photons by a magnification factor and then using this number in the simulation, or by running multiple events in a single run. In this particular instance, approximately 600 photons are generated during a single event. The results of both methods are presented in Fig.~\ref{fig:speed-up}, indicating that \opticks simulation is approximately 10 times faster in the first case and 5 times faster in the second case than pure \geant simulation. It should be noted that the computation times reported only pertain to the duration taken for optical photon generation and propagation and do not take into account the time required for geometry translation.

\subsection{Full \richone geometry}
\label{subsec:full_rich}
\begin{figure}[!ht]
    \begin{center}
    \includegraphics[width=0.99\columnwidth]{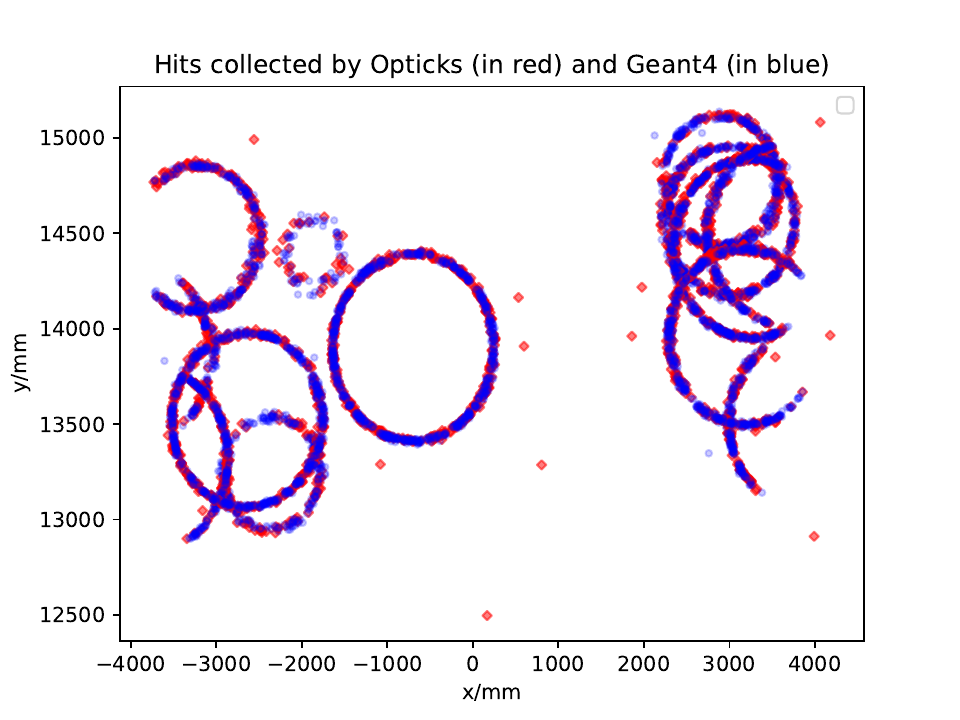}
    \end{center}
    \caption{The positions of the hits on the MaPMTs' photocathode in the $x$-$y$ plane, obtained from the simulation results of \geant (in blue) and \opticks (in red) for the \richone geometry. The particle types, directions and initial energies of the events are generated randomly.}
    \label{fig:hits_full_RICH1}
    \vspace{-2em}
\end{figure}

Upon verifying the consistency between \geant and \opticks simulation results and confirming \opticks' acceleration capabilities, a simulation of a single event through the complete \richone geometry is carried out. Figure~\ref{fig:hits_full_RICH1} displays the hits collected on the PMTs' photocathode, resulting from random charged particles passing through the \richone detector. 

From the collected hits within both \opticks and \geant, it is clear that \opticks is also well suited for the use with more than one input particle per processed event, as multiple Cherenkov rings are visible. 

\section{Discussion}
\subsection{Physics comparison of \opticks with and \geant}
\label{subsec:phys_Results}
As seen in Figure~\ref{fig:hits_full_RICH1}, the overlap of rings of multiple events shows the correct conversion of the geometry to \opticks. However, the difference between the distributions, illustrated by the shift in the mean number of hits in Figure~\ref{fig:stat_500events} and in the areas outside of the rings in Figure~\ref{fig:hits_full_RICH1}, illustrates the need for automatic porting of user defined physics processes to the \opticks framework in an automatic fashion. Regardless, the work shows that irrespective of occupancy, the core physics results are consistent. Finally, the linear scaling of results with number of events in Figure~\ref{fig:speed-up} shows optimisation of the use of GPU resources.

\subsection{Integration of \opticks with \lhcb simulation framework}

\label{subsec:integration}
Integrating \opticks with the \lhcb simulation framework\cite{Clemencic_2011,Siddi:2704513,Mazurek:2022tlu}, presents a considerable challenge, primarily due to the following two aspects:

First, \opticks relies on numerous external packages, including \geant, XercesC, CLHEP, and Boost, as well as other related libraries. These programs are all required to be specific versions for use in distributed computing resources. Although \opticks offers an automated command (\texttt{opticks-full}) to install these external packages, various issues may arise when attempting to install \opticks on a new machine. To minimise possible conflicts, a bash script is developed to update, build, and test \opticks, based on the \lhcb Nightly Build System~\cite{Clemencic:1507655}. This system is a collection of tools and scripts that automate the build and test tasks for \lhcb software, making the installation of \opticks significantly easier and providing the necessary foundation for integrating \opticks with the \lhcb simulation framework. It can also provide key insights into compatibility with future software versions.

Second, to deploy \opticks for use in simulation workloads, an interface between the \lhcb core simulation framework and \opticks needs to be developed. This necessitates major tagged releases of \opticks and the use of consistent versions of the underlying software across all software projects, highlighted in the previous paragraph, with the compatibility between all used versions to be explicitly checked. In the context of the \lhcb detector, the optical photon simulations in the \rich detectors represent one where \opticks would be invoked. The challenge lies in translating the \rich detectors from the detector geometry source provided by \gauss into the \opticks geometry buffers, interfacing other \gauss features such as algorithms, tools, and services with \opticks, loading external physics lists, and simulating the detector response. These issues within the integration process require further investigation and effort. Additionally, the detector construction relies heavily on boolean volumes, leading to an extremely unbalanced binary search tree. The balancing of this tree requires either the complete re-design of the geometry or other tools that can automatically balance the tree to be introduced. Another possible solution is the export of only the specific geometry elements for consideration for optical photon workloads by first exporting only these parts to a parallel geometry within \geant, then exported to \opticks via GDML or otherwise.

\subsection{Migration of \opticks to \optix 7.0}
\label{subsec:OptiX7.0}
\begin{figure}[!ht]
    \begin{center}
    \includegraphics[width=0.99\columnwidth]{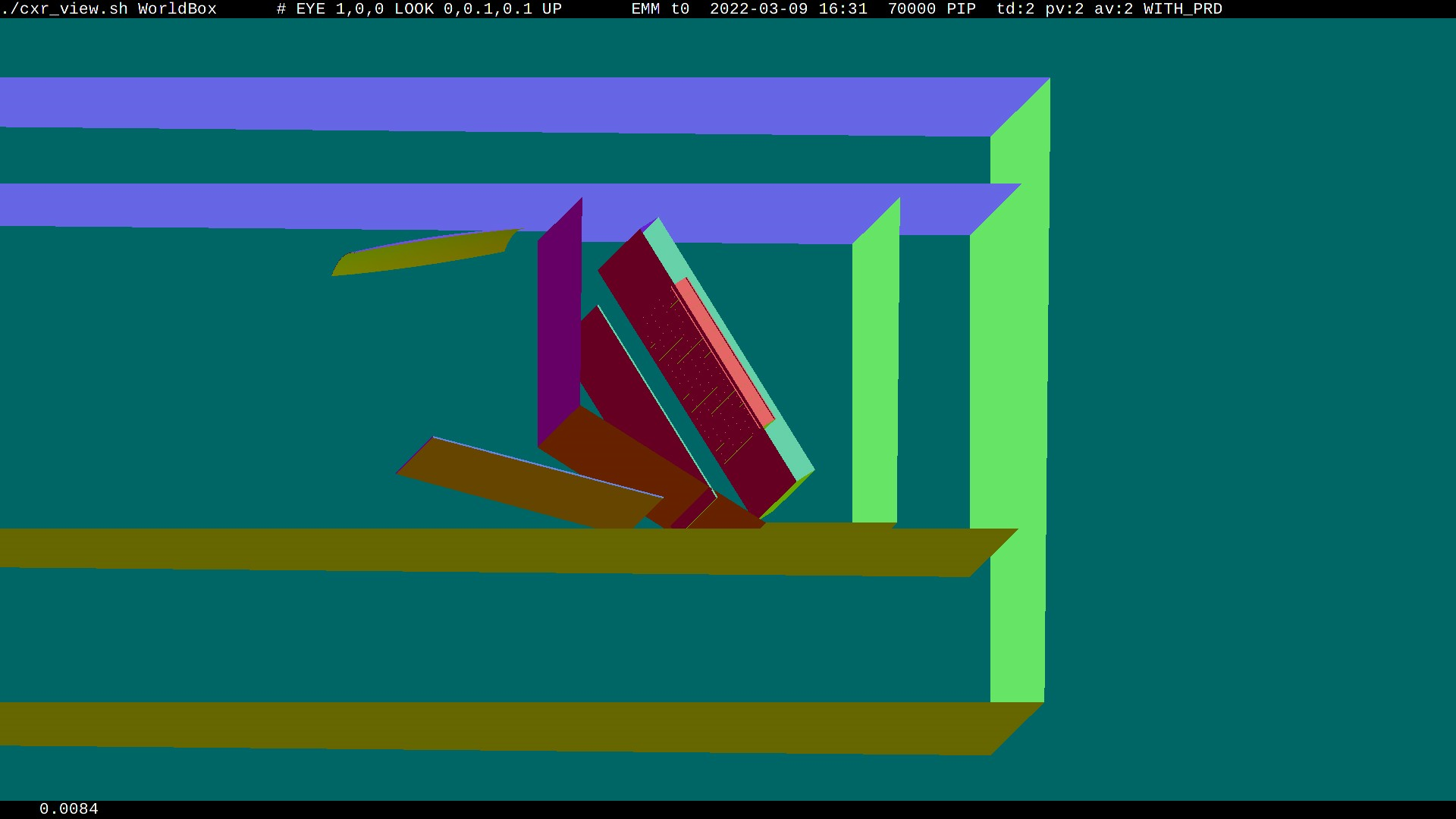}
    \includegraphics[width=0.99\columnwidth]{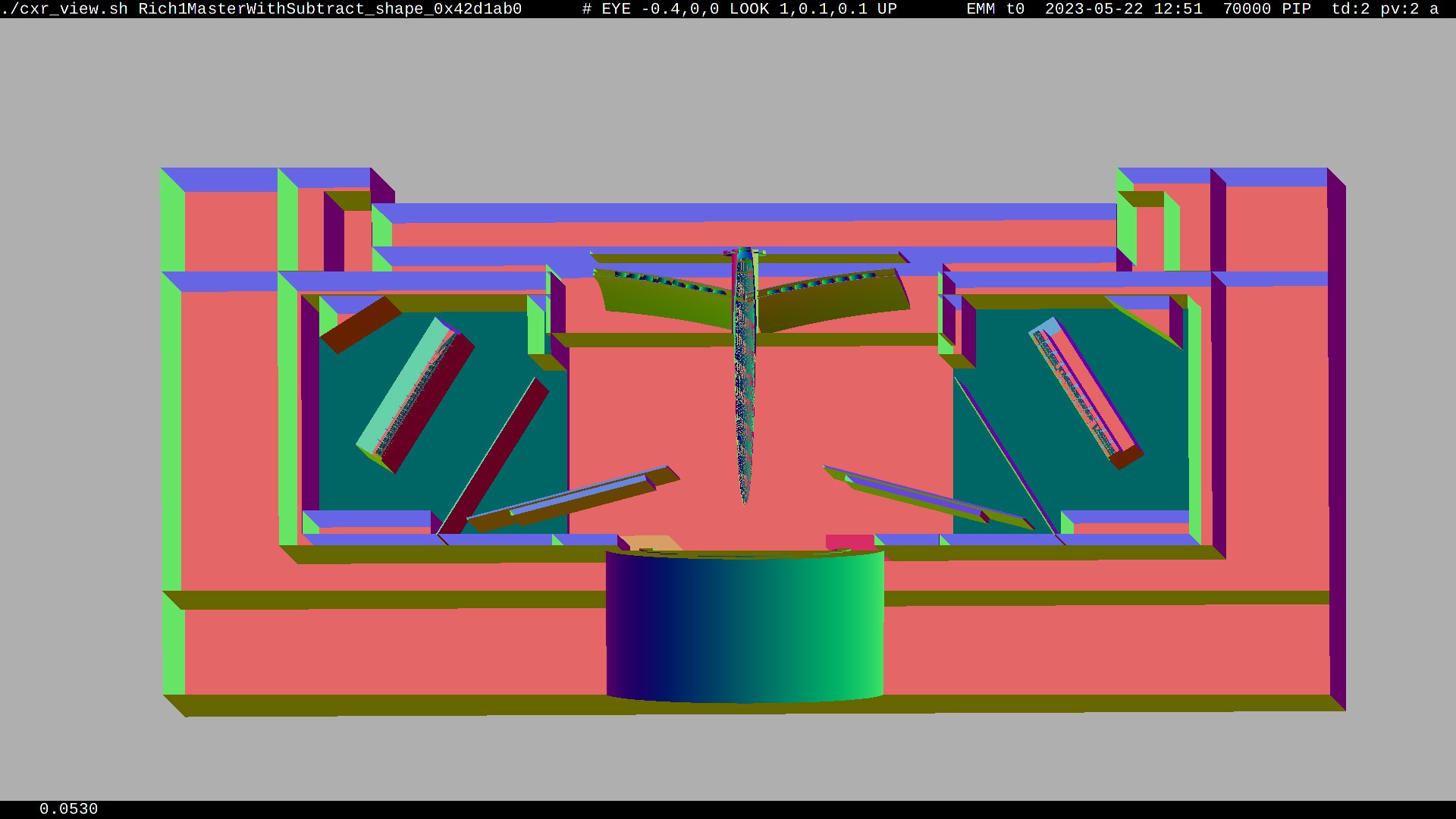}
    \end{center}
    \caption{Renders of the simplified \richone (top) and full \richone (bottom) geometries utilising NVIDIA \optix 7. The geometry is modelled using a shared CPU/GPU geometry model, known as the \texttt{CSG} geometry model, which is designed to operate with the NVIDIA \optix 7 API. The geometry is converted using the \texttt{CSG\_GGeo} class and rendered with the assistance of the \texttt{CSGOptiX} class.}
    \label{fig:OptiX7_render}
    \vspace{-2em}
\end{figure}

The code of \opticks is currently in the process of transitioning to the new NVIDIA \optix 7 API\cite{opticks_chep_2023}, and as a result, all the GPU code is currently being restructured. The \opticks \texttt{GGeo} geometry structure will be replaced by the \texttt{CSG} geometry model, and the interface between \geant code and the \opticks GPU context will be managed by the \texttt{G4CXOpticks} class. Figure~\ref{fig:OptiX7_render} presents the renderings of the simplified \richone and full \richone geometries using the \optix 7 API.

At time of study, the full suite of physics processes was unavailable for \optix 7, but the rendering capabilities were present, allowing only the report of rendering results within this paper.
While this shows the possibility of translations of geometries with the same framework but an updated interface, these geometries will still have unbalanced binary trees, meaning experiments wishing to use this will need to either optimise their geometries themselves or incur non-optimal performance. Automatic tools for the detection of such unbalanced trees may be possible, aiding users in their use of \opticks.

\section{Conclusion}
\label{sec:conclusion}
The integration of \opticks, the GPU-accelerated optical photon simulation package, with the \lhcb detector simulation framework has shown promising results in terms of efficiency improvement and consistency with \geant simulations. The hybrid workflow, which combines \geant and \opticks, has demonstrated the feasibility of offloading the CPU optical photon simulation to GPUs, improving the overall performance of the simulation process. The performance characteristics of the \texttt{Rich\_Simplified} program provided valuable insights for accelerating optical photon simulations in the \lhcb detector, particularly in the \rich subsystem, contributing to more efficient data processing and analysis. We also highlight difficulties encountered and enumerate future steps necessary for inclusion in full production environments such as those utilised within LHCb.

\section*{Acknowledgements}
%
%
\noindent We thank S. Blyth for extensive discussions and support with the Opticks package. We thank M. Kabiri Chimeh the organisers of the UK National GPU Hackathon 2022 for support. We thank R. Hohl and M. Malenta, our project mentors during the 2022 UK National GPU Hackathon, for extensive discussions and support. We also thank the LHCb online team for use of computing resources related to this work. We thank C. Fitzpatrick and M. Kreps for comments on early versions of this paper.

This work was supported by UKRI/STFC under grant agreements ST/S000925/1, ST/W000601/1, and ST/V002546/1; by the Royal Society (United Kingdom) under grant agreements DH160214 and RGF$\backslash$EA$\backslash$201014.




\bibliographystyle{LHCb}
\bibliography{main,standard,LHCb-DP}

\ifx\mcitethebibliography\mciteundefinedmacro
\PackageError{LHCb.bst}{mciteplus.sty has not been loaded}
{This bibstyle requires the use of the mciteplus package.}\fi
\providecommand{\href}[2]{#2}
\begin{mcitethebibliography}{10}
\mciteSetBstSublistMode{n}
\mciteSetBstMaxWidthForm{subitem}{\alph{mcitesubitemcount})}
\mciteSetBstSublistLabelBeginEnd{\mcitemaxwidthsubitemform\space}
{\relax}{\relax}

\bibitem{LHCb_2008}
LHCb collaboration, A.~A.~A. Jr {\em et~al.},
  \ifthenelse{\boolean{articletitles}}{\emph{{The LHCb Detector at the LHC}},
  }{}\href{https://doi.org/10.1088/1748-0221/3/08/S08005}{Journal of
  Instrumentation \textbf{3} (2008) S08005}\relax
\mciteBstWouldAddEndPuncttrue
\mciteSetBstMidEndSepPunct{\mcitedefaultmidpunct}
{\mcitedefaultendpunct}{\mcitedefaultseppunct}\relax
\EndOfBibitem
\bibitem{Mathe_2015}
Z.~Mathe, A.~C. Ramo, N.~Lazovsky, and F.~Stagni,
  \ifthenelse{\boolean{articletitles}}{\emph{The {DIRAC} web portal 2.0},
  }{}\href{https://doi.org/10.1088/1742-6596/664/6/062039}{Journal of Physics:
  Conference Series \textbf{664} (2015) 062039}\relax
\mciteBstWouldAddEndPuncttrue
\mciteSetBstMidEndSepPunct{\mcitedefaultmidpunct}
{\mcitedefaultendpunct}{\mcitedefaultseppunct}\relax
\EndOfBibitem
\bibitem{Stagni_2017}
F.~Stagni {\em et~al.}, \ifthenelse{\boolean{articletitles}}{\emph{Dirac in
  large particle physics experiments},
  }{}\href{https://doi.org/10.1088/1742-6596/898/9/092020}{Journal of Physics:
  Conference Series \textbf{898} (2017) 092020}\relax
\mciteBstWouldAddEndPuncttrue
\mciteSetBstMidEndSepPunct{\mcitedefaultmidpunct}
{\mcitedefaultendpunct}{\mcitedefaultseppunct}\relax
\EndOfBibitem
\bibitem{Agostinelli:2002hh}
Geant4 collaboration, S.~Agostinelli {\em et~al.},
  \ifthenelse{\boolean{articletitles}}{\emph{{Geant4: A simulation toolkit}},
  }{}\href{https://doi.org/10.1016/S0168-9002(03)01368-8}{Nucl.\ Instrum.\
  Meth.\  \textbf{A506} (2003) 250}\relax
\mciteBstWouldAddEndPuncttrue
\mciteSetBstMidEndSepPunct{\mcitedefaultmidpunct}
{\mcitedefaultendpunct}{\mcitedefaultseppunct}\relax
\EndOfBibitem
\bibitem{Allison:2006ve}
Geant4 collaboration, J.~Allison {\em et~al.},
  \ifthenelse{\boolean{articletitles}}{\emph{{Geant4 developments and
  applications}}, }{}\href{https://doi.org/10.1109/TNS.2006.869826}{IEEE
  Trans.\ Nucl.\ Sci.\  \textbf{53} (2006) 270}\relax
\mciteBstWouldAddEndPuncttrue
\mciteSetBstMidEndSepPunct{\mcitedefaultmidpunct}
{\mcitedefaultendpunct}{\mcitedefaultseppunct}\relax
\EndOfBibitem
\bibitem{LHCb-DP-2018-004}
D.~M{\"u}ller, M.~Clemencic, G.~Corti, and M.~Gersabeck,
  \ifthenelse{\boolean{articletitles}}{\emph{{ReDecay: A novel approach to
  speed up the simulation at LHCb}},
  }{}\href{https://doi.org/10.1140/epjc/s10052-018-6469-6}{Eur.\ Phys.\ J.\
  \textbf{C78} (2018) 1009},
  \href{http://arxiv.org/abs/1810.10362}{{\normalfont\ttfamily
  arXiv:1810.10362}}\relax
\mciteBstWouldAddEndPuncttrue
\mciteSetBstMidEndSepPunct{\mcitedefaultmidpunct}
{\mcitedefaultendpunct}{\mcitedefaultseppunct}\relax
\EndOfBibitem
\bibitem{b_coutri_sim_talk}
B.~Coutrier {\em et~al.}, \ifthenelse{\boolean{articletitles}}{\emph{{Ensuring
  simulation quality in the LHCb Experiment}}, }{}
  \url{https://indico.jlab.org/event/459/contributions/11525/}, 2023\relax
\mciteBstWouldAddEndPuncttrue
\mciteSetBstMidEndSepPunct{\mcitedefaultmidpunct}
{\mcitedefaultendpunct}{\mcitedefaultseppunct}\relax
\EndOfBibitem
\bibitem{LHCb-PROC-2011-006}
M.~Clemencic {\em et~al.}, \ifthenelse{\boolean{articletitles}}{\emph{{The
  \lhcb simulation application, Gauss: Design, evolution and experience}},
  }{}\href{https://doi.org/10.1088/1742-6596/331/3/032023}{J.\ Phys.\ Conf.\
  Ser.\  \textbf{331} (2011) 032023}\relax
\mciteBstWouldAddEndPuncttrue
\mciteSetBstMidEndSepPunct{\mcitedefaultmidpunct}
{\mcitedefaultendpunct}{\mcitedefaultseppunct}\relax
\EndOfBibitem
\bibitem{doecode_94866}
S.~R. Johnson {\em et~al.},
  \ifthenelse{\boolean{articletitles}}{\emph{Celeritas}, }{} [Computer
  Software] \url{https://doi.org/10.11578/dc.20221011.1}, 2022.
\newblock
  doi:~\href{https://doi.org/10.11578/dc.20221011.1}{10.11578/dc.20221011.1}\relax
\mciteBstWouldAddEndPuncttrue
\mciteSetBstMidEndSepPunct{\mcitedefaultmidpunct}
{\mcitedefaultendpunct}{\mcitedefaultseppunct}\relax
\EndOfBibitem
\bibitem{Amadio_2023}
G.~Amadio {\em et~al.}, \ifthenelse{\boolean{articletitles}}{\emph{{Offloading
  electromagnetic shower transport to GPUs}},
  }{}\href{https://doi.org/10.1088/1742-6596/2438/1/012055}{Journal of Physics:
  Conference Series \textbf{2438} (2023) 012055}\relax
\mciteBstWouldAddEndPuncttrue
\mciteSetBstMidEndSepPunct{\mcitedefaultmidpunct}
{\mcitedefaultendpunct}{\mcitedefaultseppunct}\relax
\EndOfBibitem
\bibitem{refId0}
{Blyth, Simon}, \ifthenelse{\boolean{articletitles}}{\emph{{Integration of JUNO
  simulation framework with Opticks: GPU accelerated optical propagation via
  NVIDIA OptiX$^\mathrm{TM}$}},
  }{}\href{https://doi.org/10.1051/epjconf/202125103009}{EPJ Web Conf.\
  \textbf{251} (2021) 03009}\relax
\mciteBstWouldAddEndPuncttrue
\mciteSetBstMidEndSepPunct{\mcitedefaultmidpunct}
{\mcitedefaultendpunct}{\mcitedefaultseppunct}\relax
\EndOfBibitem
\bibitem{An_2016}
F.~P. An {\em et~al.}, \ifthenelse{\boolean{articletitles}}{\emph{{The detector
  system of the Daya Bay reactor neutrino experiment}},
  }{}\href{https://doi.org/10.1016/j.nima.2015.11.144}{Nuclear Instruments and
  Methods in Physics Research Section A: Accelerators, Spectrometers, Detectors
  and Associated Equipment \textbf{811} (2016) 133}\relax
\mciteBstWouldAddEndPuncttrue
\mciteSetBstMidEndSepPunct{\mcitedefaultmidpunct}
{\mcitedefaultendpunct}{\mcitedefaultseppunct}\relax
\EndOfBibitem
\bibitem{JUNO_2016}
F.~An {\em et~al.}, \ifthenelse{\boolean{articletitles}}{\emph{Neutrino physics
  with {JUNO}}, }{}\href{https://doi.org/10.1088/0954-3899/43/3/030401}{Journal
  of Physics G: Nuclear and Particle Physics \textbf{43} (2016) 030401}\relax
\mciteBstWouldAddEndPuncttrue
\mciteSetBstMidEndSepPunct{\mcitedefaultmidpunct}
{\mcitedefaultendpunct}{\mcitedefaultseppunct}\relax
\EndOfBibitem
\bibitem{Parker10OptiX}
S.~G. Parker {\em et~al.}, \ifthenelse{\boolean{articletitles}}{\emph{{OptiX: A
  general purpose ray tracing engine}},
  }{}\href{https://doi.org/10.1145/1778765.1778803}{ACM Trans.\ Graph.\
  \textbf{29} (2010) }\relax
\mciteBstWouldAddEndPuncttrue
\mciteSetBstMidEndSepPunct{\mcitedefaultmidpunct}
{\mcitedefaultendpunct}{\mcitedefaultseppunct}\relax
\EndOfBibitem
\bibitem{10.1145/1365490.1365500}
J.~Nickolls, I.~Buck, M.~Garland, and K.~Skadron,
  \ifthenelse{\boolean{articletitles}}{\emph{{Scalable Parallel Programming
  with {CUDA}: Is {CUDA} the Parallel Programming Model That Application
  Developers Have Been Waiting For?}},
  }{}\href{https://doi.org/10.1145/1365490.1365500}{Queue \textbf{6} (2008)
  40–53}\relax
\mciteBstWouldAddEndPuncttrue
\mciteSetBstMidEndSepPunct{\mcitedefaultmidpunct}
{\mcitedefaultendpunct}{\mcitedefaultseppunct}\relax
\EndOfBibitem
\bibitem{10.1145/1572769.1572771}
M.~Stich, H.~Friedrich, and A.~Dietrich,
  \ifthenelse{\boolean{articletitles}}{\emph{Spatial splits in bounding volume
  hierarchies}, }{} in {\em Proceedings of the Conference on High Performance
  Graphics 2009}, \href{https://doi.org/10.1145/1572769.1572771}{ HPG '09, (New
  York, NY, USA), 7–13, Association for Computing Machinery, 2009}\relax
\mciteBstWouldAddEndPuncttrue
\mciteSetBstMidEndSepPunct{\mcitedefaultmidpunct}
{\mcitedefaultendpunct}{\mcitedefaultseppunct}\relax
\EndOfBibitem
\bibitem{opticks_chep_2023}
S.~Blyth and T.~Lin, \ifthenelse{\boolean{articletitles}}{\emph{Opticks : {GPU}
  optical photon simulation using {NVIDIA} {OptiX} 7 and {NVIDIA} {CUDA}}, }{}
  \url{https://indico.jlab.org/event/459/contributions/11811/}, 2023\relax
\mciteBstWouldAddEndPuncttrue
\mciteSetBstMidEndSepPunct{\mcitedefaultmidpunct}
{\mcitedefaultendpunct}{\mcitedefaultseppunct}\relax
\EndOfBibitem
\bibitem{Amato:494263}
LHCb collaboration, S.~Amato {\em et~al.},
  \ifthenelse{\boolean{articletitles}}{\emph{{LHCb RICH: Technical Design
  Report}}, }{} \url{https://cds.cern.ch/record/494263}, Geneva, 2000\relax
\mciteBstWouldAddEndPuncttrue
\mciteSetBstMidEndSepPunct{\mcitedefaultmidpunct}
{\mcitedefaultendpunct}{\mcitedefaultseppunct}\relax
\EndOfBibitem
\bibitem{Collaboration:1624074}
LHCb collaboration,  \ifthenelse{\boolean{articletitles}}{\emph{{LHCb PID
  Upgrade Technical Design Report}}}{},
  \href{https://cds.cern.ch/record/1624074}{CERN-LHCC-2013-022, LHCB-TDR-014},
  2013\relax
\mciteBstWouldAddEndPuncttrue
\mciteSetBstMidEndSepPunct{\mcitedefaultmidpunct}
{\mcitedefaultendpunct}{\mcitedefaultseppunct}\relax
\EndOfBibitem
\bibitem{LHCb:2023hlw}
LHCb collaboration, R.~Aaij {\em et~al.},
  \ifthenelse{\boolean{articletitles}}{\emph{{The LHCb upgrade I}},
  }{}\href{http://arxiv.org/abs/2305.10515}{{\normalfont\ttfamily
  arXiv:2305.10515}}\relax
\mciteBstWouldAddEndPuncttrue
\mciteSetBstMidEndSepPunct{\mcitedefaultmidpunct}
{\mcitedefaultendpunct}{\mcitedefaultseppunct}\relax
\EndOfBibitem
\bibitem{nvidia_nsight}
{NVIDIA Corporation}, \ifthenelse{\boolean{articletitles}}{\emph{{NVIDIA}
  {Nsight} systems}, }{} \url{https://developer.nvidia.com/nsight-systems},
  2023\relax
\mciteBstWouldAddEndPuncttrue
\mciteSetBstMidEndSepPunct{\mcitedefaultmidpunct}
{\mcitedefaultendpunct}{\mcitedefaultseppunct}\relax
\EndOfBibitem
\bibitem{Clemencic_2011}
M.~Clemencic {\em et~al.}, \ifthenelse{\boolean{articletitles}}{\emph{{The LHCb
  simulation application, Gauss: design, evolution and experience}},
  }{}\href{https://doi.org/10.1088/1742-6596/331/3/032023}{Journal of Physics:
  Conference Series \textbf{331} (2011) 032023}\relax
\mciteBstWouldAddEndPuncttrue
\mciteSetBstMidEndSepPunct{\mcitedefaultmidpunct}
{\mcitedefaultendpunct}{\mcitedefaultseppunct}\relax
\EndOfBibitem
\bibitem{Siddi:2704513}
B.~G. Siddi and D.~Muller, \ifthenelse{\boolean{articletitles}}{\emph{{Gaussino
  - a Gaudi-Based Core Simulation Framework}},
  }{}doi:~\href{https://doi.org/10.1109/NSS/MIC42101.2019.9060074}{10.1109/NSS/MIC42101.2019.9060074}\relax
\mciteBstWouldAddEndPuncttrue
\mciteSetBstMidEndSepPunct{\mcitedefaultmidpunct}
{\mcitedefaultendpunct}{\mcitedefaultseppunct}\relax
\EndOfBibitem
\bibitem{Mazurek:2022tlu}
M.~Mazurek, M.~Clemencic, and G.~Corti,
  \ifthenelse{\boolean{articletitles}}{\emph{{Gauss and Gaussino: the LHCb
  simulation software and its new experiment agnostic core framework}},
  }{}\href{https://doi.org/10.22323/1.414.0225}{PoS \textbf{ICHEP2022} (2022)
  225}\relax
\mciteBstWouldAddEndPuncttrue
\mciteSetBstMidEndSepPunct{\mcitedefaultmidpunct}
{\mcitedefaultendpunct}{\mcitedefaultseppunct}\relax
\EndOfBibitem
\bibitem{Clemencic:1507655}
M.~Clemencic and B.~Couturier,  \ifthenelse{\boolean{articletitles}}{\emph{{A
  New Nightly Build System for LHCb}}}{},
  \href{https://cds.cern.ch/record/1507655}{LHCb-INT-2013-006,
  CERN-LHCb-INT-2013-006}, CERN, Geneva, 2013\relax
\mciteBstWouldAddEndPuncttrue
\mciteSetBstMidEndSepPunct{\mcitedefaultmidpunct}
{\mcitedefaultendpunct}{\mcitedefaultseppunct}\relax
\EndOfBibitem
\end{mcitethebibliography}

\end{document}